\documentstyle[11pt]{article}
\begin{document}
\pagestyle{plain}
\huge
\title{\bf New paradox in the special theory of relativity generated
 by the string dynamics}
\large
\author{Miroslav Pardy\\[7mm]
Department of Physical Electronics \\
and\\
Laboratory of Plasma physics\\[5mm]
Masaryk University \\
Kotl\'{a}\v{r}sk\'{a} 2, 611 37 Brno, Czech Republic\\
e-mail:pamir@physics.muni.cz}
\date{\today}
\maketitle
\vspace{5mm}

\begin{abstract}
It is proved that the definition of sinultaneity by Einstein leads to the
paradox motion of the string from the viewpoint of the observer in
the inertial system $S'$ moving with velocity $v$ with regard to the inertial
system $S$.
\end{abstract}

\vspace{3mm}


\vspace{7mm}

Einstein proved in his well known article and book  (Einstein, 1905; 1919;
1922) that the simultaneity in the inertial system $S'$ moving with velocity 
$v$ with regard to the inertial system $S$ is broken. Einstein writes
(Einstein, 1905):

{\it So we see that we cannot attach any absolute signification to the
concept of simultaneity, but that two events which, viewed from a system of 
co-ordinates, are simultaneous, can no longer be looked upon as 
simultaneous events when envisaged from a system which is in motion 
relatively to that system.}

Let us show that the Einstein realization of simultaneity leads
to the paradox if we consider the dynamics of the elastic string. 

So, let the string in the system $S$ with the equilibrium length $l_{o}$
is elongated to the length $l$, $l  > l_{o}$, and it is fixed et the
ends. At time $t= 0$ the ends of he string are released. The
string motion is described by the wave equation 

$$\frac{1}{a^{2}}\frac{\partial^{2} u(x,t)}{\partial t^{2}} - 
\frac{\partial^{2} u(x,t)}{\partial x^{2}} = 0 \eqno(1)$$ 
including the initial conditions

$$u(x, 0) = Ax; \quad \dot u(x,0) = 0 \eqno(2)$$
and the boundary conditions expressing the fact that the end of the
string are free from time $t > 0$. Here $a \neq c$, $c$ being the 
velocity of light.

Without mathematics we know that the center of mass of string is at
the rest. This result can be immediately confirmed using the rubber
string with the fixation of ends by fingers. In other words, by the
string-finger experiment. The experiment with the rubber string is
very 
simple and can be performed everywhere.

The situation in the system $S'$ moving with the velocity $v$ with
regard to the system $S$ is different because the releasing of the
ends of the rubber string is not simultaneous according the Einstein
definition of simultaneity. 
It means that the motion of the string is a such that
the center of mass changes its velocity. We also can verify the change
of velocity of the center of mass by the rubber string experiment
with the noninstantaneous releasing of the ends of the string.

So, we have a paradox. The center of mass of he string in system $S$ does
not change its state of motion, while in the system $S'$ it does. This
paradox  is not involved 
in the collection of paradoxes of  relativity (Goldblatt, 1972;
Terletzkii, 1966) and in the relativistic  paradoxes in American Journal of
Physics. To our knowledge, this paradox is not involved in any
monograph of the string theory.

There is no doubt that solution of this paradox will refresh the
interest in the theory of real strings described by the equations of
mathematical physics.

Let us remark only that the paradox can be resolved by observing that
the 
wave equation (1) is not relativistically invariant with regard to the 
Lorentz transformation

$$x' = \gamma(x -vt), \quad t' = \gamma(t - (v/c^{2})x); 
\quad \gamma = \frac{1}{\sqrt{1 - v^{2}/c^{2}}}. \eqno(3)$$

It means that it is necessary to transform equation (1) to the system
$S'$ and 
then to solve the motion of strings with the broken boundary
condition. To 
our knowledge the solution of this problem was not published till this time. 

The resolution of the paradox with the broken simultaneity is
important not 
only
from the viewpoint of relativity  but from the viewpoint of pedagogical
thinking and from the viewpoint of the methodology of science. The
discussion of such problems as simultaneity and string motion in the 
different inertial systems is important for  the logical
foundation of science and it means that every elementary new information,
if true, is of great influence on science.

Let us remark that we get another paradox if we consider equation (1) with the 
boundary conditions

$$u(0,t) = 0; \quad u(l,t) = 0 \eqno(4)$$
and the initial conditions in the general form

$$u(x, 0) = f(x); \quad \dot u(x,0) = g(x) \eqno(5)$$
where $f,g$ are arbitrary functions.

It is well known that the general solution is of the form

$$u = \varphi(t + x/a) + \psi(t - x/a) \eqno(6)$$

However, the solution of such vibration of the string is also of the
form (which was known also to Daniel Bernoulli and Leonhard Euler):

$$u = \sum_{k=1}^{\infty} \left(a_{k}\cos\frac{k\pi at}{l_{0}} + 
b_{k}\sin\frac{k\pi at}{l_{0}}\right)\sin\frac{k\pi x}{l_{0}}
\eqno(7)$$

If we transform the solution (6) to the system $S'$, then we transform
only coordinate $x$ and time $t$. On the other hand, if we transform the 
solution (7), then we must also respect the Lorentz contraction of
$l_{0}$. And we are not sure, a priori, that the two results will be
identical. And this is the paradox.

\vspace{1cm}

{\bf References}.

\vspace{1cm} 

\noindent
Einstein, A. (1905). Zur Elektrodynamik bewegter K{\"o}rper, Annalen der
Physik, {\bf 17}. See also {\it The Principle of Relativity}, 
published in 1923 by Methuen and Company, Ltd. of London. Most of the 
papers in that collection are English translations by W. Perrett and 
G.B. Jeffery from the German {\it Das Relatitivit{\"a}tsprinzip}, 4th ed., published 
in 1922 by Teubner.  \\[5mm]
Einstein, A. (1919). {\it \"Uber die spezielle und die allgemeine
relativit{\"a}ts Theorie}, Vierte Auflage, Vieweg {\&} Sohn, 
Braunschweig.; chapter 9.\\[5mm]
Goldenblat, I. I. (1972). {\it The time paradoxes in the special
theory of relativity},  (NAUKA, Moscow), (in Russian). \\[5mm]
Terletzkii, Ya. P. (1966). {\it The paradoxes of the special theory
of relativity}, (NAUKA, Moscow), (in Russian).

\end{document}